# Specific heat of robust $Nb_2PdS_5$ superconductor

**Reena Goyal, Brajesh Tiwari, Rajveer Jha and V. P. S. Awana[*]**

CSIR-National Physical Laboratory, Dr. K. S. Krishnan Marg, New Delhi-110012, India

We report specific heat under different magnetic fields for recently discovered quasi-one dimensional $Nb_2PdS_5$ superconductor. The studied compound is superconducting below 6 K. $Nb_2PdS_5$ is quite robust against magnetic field with $dH_c/dT$ of -42 kOe/K. The estimated upper critical field [$H_{c2}(0)$] is 190 kOe, clearly surpassing the Pauli-paramagnetic limit of $1.84 T_c$. Low temperature heat capacity in superconducting state of $Nb_2PdS_5$ under different magnetic fields showed s-wave superconductivity with two different gaps. Two quasi-linear slopes in Somerfield-coefficient ($\gamma$) as a function of applied magnetic field and two band behavior of the electronic heat capacity demonstrate that $Nb_2PdS_5$ is a multiband superconductor in weak coupling limit with $\Delta C/\gamma T_c$=0.9.

**Introduction:** Superconductivity has always been fascinating to the researchers from both experimental and theoretical condensed matter physics communities. In this regards, very recent observation of superconductivity below 6K in quasi-one dimensional $PdS_2$ chains based $Nb_2PdS_5$ has attracted a lot of attention due to its large upper critical field, exceeding the Pauli paramagnetic limit [1-6]. The estimated upper critical field [$H_{c2}(0)$] for $Nb_2PdS_5$ of around 200 kOe is simply outside Pauli paramagnetic limit of $1.84 T_c$ [1-4]. $Nb_2PdS_5$ crystallizes in centro-symmetric monoclinic structure within space group C2/m, and the upper critical field is remarkably large along the $PdS_2$ chain. In a very recent report, Zhou et al. have showed that spin-orbit coupling [7] along with multiband effects [1, 2, 5] may have significant impact on larger upper critical field of $Nb_2PdS_5$. The temperature dependent anisotropy in upper critical field along with first principle electronic structure calculations further substantiated the multi band nature of $Nb_2PdS_5$ superconductor [1]. The heat capacity studies on S site Se doped $Nb_2PdS_5$ showed that $Nb_2Pd(S_{1-x}Se_x)_5$ is a fully gapped superconductor within weak coupling limit [2]. The superconductivity is also observed in a similar compound $Nb_2Pd_xSe_5$ with extraordinarily large upper critical field of 350 kOe [5]. Further, multiband superconductivity is claimed for $Nb_2Pd_xSe_5$ superconductor by fitting electronic heat capacity data with two-band alfa model [5]. In any case, the clear understanding of exotic properties of $Nb_2PdS_5$ superconductor remains elusive [1-7]. Though there have been various experimental and theoretical techniques developed to study the thermodynamic properties of superconductors, the specific heat has an advantage as it probes the bulk and avoids artifacts from surface and interfaces. Though, preliminary heat capacity data were published by some of us, but we could not explain or justify the observed small value of heat capacity jump at $T_c$ [4]. To confirm this point we repeated the work and came to the point that this small heat capacity jump is intrinsic to $Nb_2PdS_5$. The sample used in current MS is newly synthesized and the physical property measurements are done a fresh. To confirm the possible two gap nature of $Nb_2PdS_5$, we did specific heat measurements at fixed low temperature under magnetic field. $C_p(H)$ is reported by some of us in ref.4 as well, but with varying T. The electronic heat capacity in superconducting regime under magnetic field can yield important information about the superconducting gap. But it is difficult to extract the same accurately from total heat capacity because of dominating phonon contributions. One way to deal with is to measure heat capacity under magnetic field at temperatures much below $T_c$ of a superconductor, preferably at $T_c/3$. In this article the T is fixed at 2K, i.e., at 1/3 of $T_c$ and field is varied in steps and Cp is measured. The careful analysis showed that $Nb_2PdS_5$ have two s-wave superconducting gap (fully gapped) as supported by two distinct linear slopes in Cp(H) at 2K, i.e., at temperature well below the $T_c$. Also, the superconducting heat capacity jump is well fitted to the two gap alfa model. In this work, the specific heat measurements have been carried out under magnetic field to determine the nature and magnitude of the quasiparticle energy gap of $Nb_2PdS_5$. It is found that $Nb_2PdS_5$ is an s-wave multiband superconductor in weak coupling limit with $\Delta C/\gamma T_c$ =0.9.

**Experimental Details:** $Nb_2PdS_5$ was synthesized via solid state reaction route. The powder of Nb, Pd and S with purity better than 99.9% were taken in stoichiometric ratio of 2:1:6 and mixed thoroughly inside argon filled glove-box before pelletized in rectangular shape under uniaxial stress of 100kg/cm$^2$. The excess Sulfur is needed to compensate the loss during the heating process due to high vapor pressure of S. We have followed same experimental procedure for synthesizing our material as described in ref. [2]. The structure and phase purity were determined with Rigaku X-ray diffractometer using CuK$_\alpha$ line of 1.54184Å. Transport and specific heat measurements were performed under applied fields on Quantum Design (*QD*) Physical Property Measurement System (*PPMS*) down to 2 K.

**Results and discussion:** The powder x-ray diffraction pattern of as synthesized $Nb_2PdS_5$ sample is shown in Figure 1. The dominant peaks are indexed according to the monoclinic structure with space group C2/m.

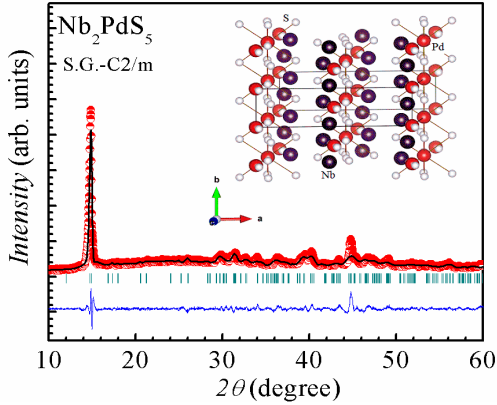

**Figure 1 (Color online) Powder x-ray diffraction pattern of $Nb_2PdS_5$. Inset shows the crystal structure**

The detailed structural analysis and their structural parameters are almost similar to our previous work on $Nb_2PdS_5$ [4]. The estimated lattice parameters are a=12.21 Å, b=3.27 Å, c=15.23 Å and β=103.44°. Inset shows the crystal structure, where chain of $PdS_2$ can be seen along crystallographic *b*-axis.

Figure 2 displays temperature dependence of electrical resistivity $\rho(T)$ of $Nb_2PdS_5$ with and without applied magnetic field. It is clear from Figure 2, that resistivity of sample increases with temperature i.e., a metallic normal state and a superconducting transition is observed with onset $T_c$~6.6 K and $T_c(\rho=0)$ at 6.0 K. At low temperatures (8 K to 45 K), the resistivity is not linear with temperature, but follows $\rho=\rho_0+AT^2$, suggesting Fermi-liquid nature. Fitting the $\rho(T)$ data resulted in residual resistivity $\rho_0$=1.310 mΩ-cm and A=1.249×10$^{-4}$ mΩ-cm/K$^2$, which are in general agreement with earlier reports on same compound [1-4]. The magneto resistivity $[\rho(T)H]$ in superconducting regime is shown in inset of Figure 2. With the use of magneto resistivity curves at different fields, we estimate the upper critical field $H_{c2}(0)$ by relation $H_{c2}(0)=-0.69T_c(dH_c/dT)_{T=T_c}$. The relation is based on model developed by Werthamer, Helfand, and Hohenberg (WHH) and is true for all BCS type superconductors. In present case the experimental value of $dH_c/dT$ is -42k Oe/K, and thus the calculated $H_{c2}(0)$ is 191 kOe. Interestingly, the permitted value by Pauli paramagnetic limit is around 90 kOe by relation $H_{c2}(0)=1.84T_c$. The extraordinary upper critical field as estimated from magneto-resistivity measurements demands a detailed study on thermodynamic property and the nature of superconducting gap for $Nb_2PdS_5$.

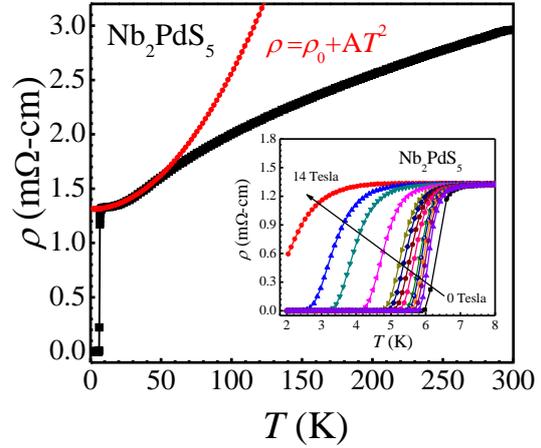

**Figure 2 (Color online) Resistivity versus temperature dependent of $Nb_2PdS_5$, Inset shows same measurement with several applied fields of 0 to 140kOe below 10K.**

Heat capacity being a bulk thermodynamic quantity is a sensitive probe at low lying quasi particles excitations, which could provide important information about the symmetry and magnitude of superconducting gap. In Figure 3, we show the variation of specific heat with temperature at different applied magnetic fields. From this Figure we can conclude that there is a transition jump from superconducting state to normal state at critical temperature of 6.0 K, which is consistent with the resistivity measurements. The superconducting transition jump is gradually suppressed and shifted to lower temperatures with increase of applied magnetic field.

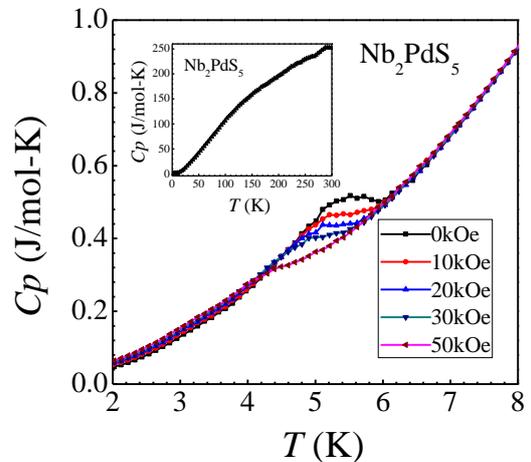

**Figure 3(Color online) Variation of specific heat with temperature at several applied fields. The inset shows specific heat without field in full range 2 to 300K.**

The full range (2-300 K) heat capacity without applied magnetic field is shown in inset of Figure 3, which is devoid of any phase transitions above $T_c$, excluding any other order/disorder phases in the studied $Nb_2PdS_5$.

Figure 4 displays variation of electronic specific heat as $C_{es}/T$ against temperature $T$. It can be observed that superconducting jump is suppressed on increasing field along with decrease in $T_c$. The normal state (above $T_c$) heat capacity at different applied fields merges, implying that the sample is free from magnetic impurities. A shallow shoulder at around 3.5 K has also been observed in superconducting state of $Nb_2PdS_5$, indicating a second superconducting gap, similar to well-known example $MgB_2$ [8]. Since the position of shoulder is not changing visibly with the applied magnetic field, we can say that it may be due to some impurity phase present in the $Nb_2PdS_5$.

It is worth noting that residual Somerfield coefficient in superconducting state (i.e. $C_{es}/T$ at 2 K) increases with external magnetic fields, suggesting that the sample has normal fractions, where superconductivity is broken by external magnetic field in spite of superconducting state. The normal state (metal) specific heat has two major contributions $C=\gamma_n T+\beta T^3$. The first one is from electronic contribution, which is linear in $T$ ($\gamma_n T$) and the second one is from phononic, which is cubic in $T$ ($\beta_n T^3$). Where, $\gamma_n$ is Somerfield coefficient and $\beta_n$ is the Somerfield-Debye expression. Inset of Figure 4 shows the plot of $C/T$ against $T^2$ in low temperature range without applied field.

The intercept gives the value $\gamma_n$ and the slope is the value of Debye constant $\beta_n$. The fitting result is shown as solid red line in inset of Figure 4, giving values of $\gamma_n$ and $\beta_n$ equal to 41.75 mJ/mol-$K^2$ and 1.14 mJ/mol-$K^4$ respectively. We estimated the Debye temperature $\Theta_D$=251.7 K, using the relation $\Theta_D=(12\pi^4 N_A z k_B/5\beta_n)^{1/3}$, where $N_A$(=$6.022\times10^{23}$ $mol^{-1}$) is the Avogadro's constant, z(=9) is the number of atoms per formula unit and $k_B$(=$1.38\times10^{-23}$ $m^2 kg$ $s^{-2}K^{-1}$) is the Boltzmann's constant. The $\Theta_D$(251.7 K) is much higher than superconducting transition temperature of 6 K. The electronic heat capacity ($C_{es}$) in superconducting regime can yield important information about the superconducting gap [8-10]. But it is difficult to extract $C_{es}$ accurately from total heat capacity because of dominating phonon contributions. One way to deal with is to measure heat capacity under magnetic field at deep below (at least one third) temperatures of the $T_c$ of a superconductor and extract the change in $C_{es}$ or $\gamma$. This way one gets safely the $\Delta\gamma(H)$, which is nothing but $\gamma(H)-\gamma(0)$.

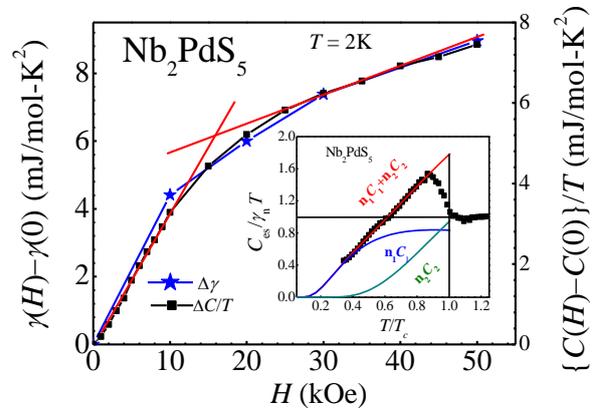

Figure 5 (Color online) Variation of normalized Sommerfeld-coefficient and specific heat against magnetic field at 2K. Inset: variation of electronic specific heat with reduced temperature $T/T_c$, where solid red curve shows two band model fitting of same measurement.

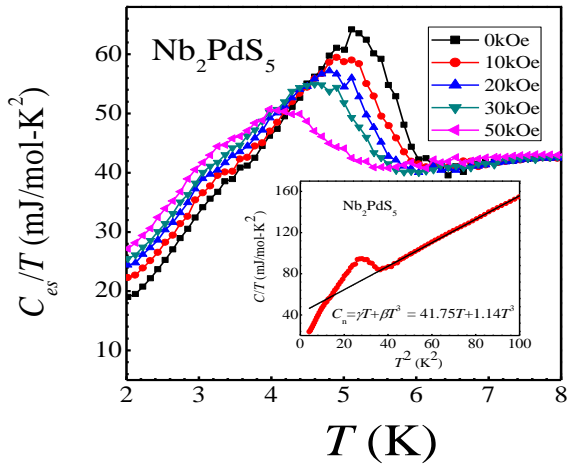

Figure 4 (Color online) Electronic specific heat measurement represented as $C_{es}/T$ vs. $T$ under various applied fields; inset shows Zero field specific heat as $C/T$ versus $T^2$. The solid red line shows linear fit to $C/T=\gamma_n+\beta T^2$.

The magnetic field dependence of obtained $\Delta\gamma(H)$ is presented in Figure 5 in mixed state of $Nb_2PdS_5$ superconductor, which is complemented though in approximation with $\Delta C/T$ at $T=2$ K. It is important to note that $\Delta\gamma(H)$ is quasi-linear with two clear slopes as a function of external field, implying the presence of two gaped superconductivity. For fully gaped s-wave superconductor, $\Delta\gamma$ varies linearly with applied magnetic field as number of vortex grow with field [10]. Here, $\Delta\gamma(H)$ behavior can be understood with help of two-band model, where $\Delta\gamma(H)$ is decomposed into two linear parts, one up to 10kOe and another from 25 kOe to 50 kOe applied field ranges. Inset of Figure 5 shows the variation of electronic specific heat ($C_{es}$) as a function of reduced temperature $T/T_c$. Electronic specific heat data have been fitted using exponential function, $C_{es}$= exp(-$\Delta$/k$_B$T). The best fit (solid red line) is obtained for two band model i.e, $C_{es}$=n$_1$$C_1$+n$_2$$C_2$, which further justifies the claim of multiband superconductivity in studied Nb$_2$PdS$_5$ superconductor. The estimated superconducting gaps are 2$\Delta_1$/k$_B$T$_c$= 6.4 for major (91%) and only 9% for 2$\Delta_1$/k$_B$T$_c$=1.9 which need to be corroborated with further low temperature measurements. The estimated jump in heat capacity at superconducting transition i.e., $\Delta C/\gamma T_c$ is 0.9, which is much lower than the one as predicted by BCS single gap superconductor but close to popular two gap superconductor MgB$_2$[8,11]. It is clear from specific heat measurements under different magnetic fields that Nb$_2$PdS$_5$ is quasi-one dimensional two gap superconductor, which need to be complimented by other experimental techniques in future.

**Conclusions:** In conclusion, we have studied the electronic and heat capacity properties of a recently discovered quasi-one dimensional superconducting compound Nb$_2$PdS$_5$ with $T_c$=6 K. The upper critical field of bulk polycrystalline sample is 191 kOe. The estimated jump in heat capacity at superconducting transition i.e., $\Delta C/\gamma T_c$ is 0.9, which is much lower than the one as predicted by BCS single gap superconductor. Low temperature heat capacity in superconducting state of Nb$_2$PdS$_5$ under different magnetic fields unambiguously confirmed the two gap nature of Nb$_2$PdS$_5$ for first time.

**Acknowledgements** Reena Goyal would like to thank UGC, for research fellowship. This research work is financially supported under *DAE-SRC* outstanding investigator award scheme on search for new superconductors.